\documentclass[showpacs,preprintnumbers,amsmath,amssymb,superscriptaddress,nofootinbib]{revtex4}
\usepackage{graphicx}% Include figure files
\usepackage{dcolumn}% Align table columns on decimal point
\usepackage{bm}% bold math
\usepackage{atlasphysics}

\begin{document}

\title{Normalizing Weak Boson Pair Production at the Large Hadron Collider}
\author{J.~M.~Campbell}
\affiliation{Department of Physics and Astronomy, University of Glasgow, Glasgow G12 8QQ, UK}
\author{E.~Castaneda-Miranda}
\affiliation{CINVESTAV-IPN, Physics Department, Mexico City 07360, Mexico}
\affiliation{Department of Physics, University of Wisconsin, Madison, WI 53706, USA}
\author{Y.~Fang}
\affiliation{Department of Physics, University of Wisconsin, Madison, WI 53706, USA}
\author{N.~Kauer}
\affiliation{Department of Physics, Royal Holloway, University of London, Egham TW20 0EX, UK}
\author{B.~Mellado}
\author{Sau Lan Wu}
\affiliation{Department of Physics, University of Wisconsin, Madison, WI 53706, USA}

%\date{\today}

\begin{abstract}
\vspace*{0.1cm}
The production of two weak bosons at the Large Hadron Collider will be one of the most important sources of SM backgrounds for final states with multiple leptons. In this paper we consider several quantities that can help normalize the production of weak boson pairs.
%%% John 
Ratios of inclusive cross-sections for production of two weak bosons and Drell-Yan are investigated and the corresponding theoretical errors are evaluated. The possibility of predicting the jet veto survival probability of $VV$ production from Drell-Yan data is also considered. Overall, the theoretical errors on all quantities remain less than $5\div 20~\%$. The dependence of these quantities on the center of mass energy of the proton-proton collision is also studied. 
%%% John 

\end{abstract}

%\keywords{}

\maketitle

\section{Introduction}
\label{sec:intro}
The physics potential of the Large Hadron Collider (LHC) is breath taking. The CMS and ATLAS experiments at the LHC are expected to shed light on the origin of mass and dark matter of the universe, and other crucial aspects involving physics beyond the Standard Model (SM)~\cite{CMSPTDR,LHCC99-14,Aad:2009wy}. The investigation of the  phase-space provided by the large center of mass energy of the proton-proton collision and the multiplicity of experimental signatures make the prospects of searches for the Higgs boson and physics beyond the SM at the LHC very exciting. 

The search for new physics in final states with multiple charged leptons~\footnote{In this paper by leptons we imply electron and muons only.} and missing transverse energy (\met) carried by neutrinos or other particles escaping detection is arguably one of the most interesting among the feasible signatures at hadron colliders. The presence of several isolated leptons reduces the contribution of expected SM processes, allowing for the exploration of physics beyond the SM in low $\met$ regions. 

The search for $WW, ZZ$ resonances is central for the observation of the Higgs boson in a wide range of mass~\cite{Keung:1984hn,Glover:1988fn,Barger:1990mn,Dittmar:1996ss,Han:1998ma,Kauer:2000hi,Anastasiou:2007mz,Anastasiou:2008ik,Grazzini:2008tf,Mellado:2007fb,Schmidt:2009kk}. Several are the motivations for searches for physics beyond the SM with these final states. The production of multi-leptons is predicted by R-parity conserving~\cite{Haber:1984rc,Gupta:2007ui,Bhattacharya:2008qu,Bhattacharya:2008dk} and violating~\cite{Barbier:2004ez} supersymmetry and nonminimal supersymmetric models~\cite{Barger:2005hb}. %%% John
The existence of massive neutrinos is a strong motivation for physics beyond the SM and in order to test their Majorana nature one has to investigate the possibility of observing them in different multi-lepton channels~\cite{Han:2006ip,Aguila:2006dx,Muhlleitner:2003me,BarShalom:2008gt,Perez:2008zc,Perez:2008ha}. %%% John
The phenomenology of the Littlest Higgs boson models with T-parity also predicts significant rates of multi-leptons~\cite{Datta:2007xy}. Furthermore, kinematic properties of the production of multi-leptons can be used to disentangle models beyond the SM~\cite{Belyaev:2008pk}.

The production of two weak bosons ($VV$) at the LHC will be one of the most important sources of SM backgrounds for multi-lepton final states. The prediction of the rates of $VV$ will be instrumental in these searches. In this paper we consider several quantities that can help normalize the production of $VV$, as suggested in~\cite{Dittmar:1997md}. These quantities will have reduced uncertainties since the error due to the integrated luminosity measurement will cancel out, and the scale and parton density uncertainties will be diminished. Ratios involving the production of $VV$ and Drell-Yan are investigated and the corresponding theoretical errors are evaluated. In Section~\ref{sec:inclusive} we investigate a number of  ratios to predict the inclusive rates of $ZZ, WW$ and $ZW$ production. In Section~\ref{sec:jvsp} we consider the possibility of predicting the jet veto survival probability of $VV$ with Drell-Yan. The latter is crucial to distinguish the multi-lepton backgrounds coming from $VV$ from those coming from the decays of top quarks.

The LHC is scheduled to deliver proton-proton collisions in 2009 and 2010 with a center of mass energy, $\sqrt{s}=10\,\tev$, after which efforts will be made to reach $\sqrt{s}=14\,\tev$~\cite{ChamonixProc}. The nominal results presented in this paper have been obtained for $\sqrt{s}=14\,\tev$, but we have also considered their dependence on lower values of $\sqrt{s}$.

\section{Inclusive Rates}
\label{sec:inclusive}

The inclusive rate of multi-leptons coming from the decay of $VV$ can be predicted by multiplying the observed rate of di-leptons coming from the Drell-Yan process by the corresponding cross-section ratios ${\sigma_{VV} \over \sigma_{Z^{(\ast)}}}$ (and ${\sigma_{VV} \over \sigma_{V'V'}}$, when appropriate) computed theoretically. The uncertainty associated with these estimates arises from experimental and theoretical errors. The experimental errors arise primarily from the uncertainty of the rate of di-leptons coming from Drell-Yan and the error on the efficiency of observing multi-leptons coming from the decay of $VV$. The dominant theoretical errors are expected to arise from the uncertainty  of the ratios ${\sigma_{VV} \over \sigma_{Z^{(\ast)}}}$ due to QCD scale variations and uncertainties in the parton densities in the proton. 
%Other theoretical uncertainties related to the weak boson(s) rapidity distributions, which can affect the geometrical acceptance of detection, are not considered here. 

The results on ${\sigma_{VV} \over \sigma_{Z^{(\ast)}}}$ (and ${\sigma_{VV} \over \sigma_{V'V'}}$) shown in this Section are based on fixed order Matrix Element (ME) computations, including the decay products of the weak bosons. Cross-sections are computed for decays involving leptons of different flavor after the application of generic kinematic cuts. The renormalization and factorization scales are set to the average mass of the weak boson. The scale-related uncertainty is obtained by varying the renormalization and factorization scales by a factor of four. Here we change both scales at the same time in equal and in opposite directions and we take the maximum deviation for the quantity considered. We believe this choice of scales yields a conservative estimate of the scale-related error.
%%% John 
The CTEQ6M~\cite{JHEP_0207_012} parton density parameterization is used, which also allows an estimate of the corresponding parton density uncertainties. We stress that results are obtained at parton level and no detector simulation is implemented. Our choice of electro-weak parameters corresponds to setting the Fermi constant, $G_F$ and the masses of the weak bosons and the top quark, $m_W$, $m_Z$, $m_t$, and then determining $\alpha_{Z}$ and $\sin^2{\theta_W}$ from them~\footnote{See http://mcfm.fnal.gov/ for more details.}.
%%% John 

\subsection{{\boldmath $ZZ$} Production}
\label{sec:zz}

%%% John 
Accurate predictions for the hadronic production of $Z$-boson pairs, including higher order QCD corrections, have been studied extensively in the literature~\cite{pr_43_3626,Mele:1990bq,Campbell:1999ah}. 
%%% John 
The production of $Z$ boson pairs through gluon-gluon fusion 
contributes at ${\cal O}(\alpha_s^2)$ relative to $q\bar{q}$ annihilation, 
but its importance is enhanced by the large gluon flux at the LHC. 
This process was first analyzed in Refs.~\cite{Dicus:1987dj,Glover:1988rg}, with leptonic decays subsequently studied for on-shell~\cite{Matsuura:1991pj} and 
off-shell~\cite{Zecher:1994kb} weak bosons. The first complete calculation of the gluon-induced loop process 
$gg \to Z^\ast(\gamma^\ast)Z^\ast(\gamma^\ast) \to \ell\bar{\ell}\ell'\bar{\ell'}$~\footnote{For simplicity we sometimes refer to this process as $gg\rightarrow ZZ$.},
allowing for arbitrary invariant masses of the $Z$ bosons and including also the 
photon contributions, was presented in Ref.~\cite{Binoth:2008pr}. 
For Higgs boson searches with masses below the $Z$-pair threshold, the virtual photon contribution to the 
$Z^\ast(\gamma^\ast)Z^\ast(\gamma^\ast)$ background cannot be neglected, since 
one of the produced $Z$ bosons will almost always be off resonance. Here we focus on the production of two lepton pairs with invariant masses close to the $Z$ mass.
%The cross-sections computed here correspond to the ME reported in Ref.~\cite{Binoth:2008pr}.

\begin{table}[t]
\caption{Cross-sections (in fb) for the $ZZ$ and $Z^\ast$ production at the LHC for different ranges of the invariant mass of the leptonic system, $m_{N\ell}$ (in $\gev$). The renormalization and factorization scales  are set to be equal to the $Z$ mass. The event selection specified in Section~\ref{sec:zz} is applied. Results are given for $\sqrt{s}=14\,\tev$. \label{tab:nominalzztoz_static}}
\begin{center}
\begin{tabular}{|c|c|c|c|c|}
\hline
$m_{N\ell}$ range &$\sigma_{q\overline{q}
\rightarrow Z^\ast}^{NLO}$ & $\sigma_{q\overline{q}
\rightarrow ZZ}^{NLO}$& $\sigma_{gg
\rightarrow ZZ}^{LO}$&  ${ \sigma_{ZZ} \over \sigma_{Z^\ast}}\cdot 10^3$ \\
\hline
%200 -  250  & 1773.7 &  7.99 & 1.182 &  5.17 \\ 
%250 -  300  &  753.2 &  3.65 & 0.530 &  5.54 \\ 
%300 -  350  &  372.4 &  1.86 & 0.246 &  5.66 \\ 
%350 -  400  &  205.7 &  1.07 & 0.131 &  5.83 \\ 
%400 -  450  &  121.0 &  0.64 & 0.082 &  5.94 \\ 
%450 -  500  &   76.0 &  0.40 & 0.055 &  6.01 \\ 
%500 -  750  &  143.9 &  0.74 & 0.114 &  5.92 \\ 
%750 - 1000  &   27.4 &  0.16 & 0.033 &  6.88 \\ 
200 -  250  &  886.8 &  4.00 & 0.591 &  5.17 \\ 
250 -  300  &  376.6 &  1.82 & 0.265 &  5.54 \\ 
300 -  350  &  186.2 &  0.93 & 0.123 &  5.66 \\ 
350 -  400  &  102.8 &  0.53 & 0.066 &  5.83 \\ 
400 -  450  &   60.5 &  0.32 & 0.041 &  5.94 \\ 
450 -  500  &   38.0 &  0.20 & 0.027 &  6.01 \\ 
500 -  750  &   71.9 &  0.37 & 0.057 &  5.92 \\ 
750 - 1000  &   13.7 &  0.08 & 0.016 &  6.88 \\ 
\hline
\end{tabular}
\end{center}
\end{table}

A first attempt to relate the production of four leptons arising from $ZZ$ production to di-lepton production from the Drell-Yan process in a proton-proton collision was reported in Ref.~\cite{Buttar:2006zd}.
%%% John 
Although the next-to-leading order (NLO) QCD corrections to the $q\overline{q}$ initiated processes were taken into account, the necessary tools for the evaluation of the contribution from the $gg\to ZZ$ diagrams referred to above were not available. These corrections are added in this paper.
%%% John 

We use the package MCFM for the computation of the cross-sections due to $q\overline{q}\to Z^\ast(\gamma^\ast)Z^\ast(\gamma^\ast)\to \ell\bar{\ell}\ell'\bar{\ell'}$ in NLO QCD, $\sigma_{q\overline{q}\rightarrow ZZ}^{NLO}$~\cite{Campbell:1999ah}.
%%% John 
The matrix elements for $gg \to Z^\ast(\gamma^\ast)Z^\ast(\gamma^\ast) \to \ell\bar{\ell}\ell'\bar{\ell'}$ are implemented in the package {\tt gg2ZZ}, a parton-level integrator and event generator~\cite{Binoth:2008pr} that we use to compute the corresponding leading order cross section, $\sigma_{gg\rightarrow ZZ}^{LO}$. The MCFM package is also used for the computation of the Drell-Yan QCD NLO cross-section $\sigma_{q\overline{q} \rightarrow Z^{(\ast)}}^{NLO}$. Since a complete NNLO QCD description of $ZZ$ production is not available, a best estimate is obtained by summing the cross-sections $\sigma_{q\overline{q}\rightarrow ZZ}^{NLO}$ and $\sigma_{gg\rightarrow ZZ}^{LO}$, treating them as independent processes. 
%%% John 
%This will lead to conservative estimates of the theoretical errors.
 
We compute the cross-sections for $ZZ$ production with generic cuts on the decay products. It is required that each of the four leptons has a transverse momentum, $p_T>20\gev$ and lies in the pseudorapidity range, $\left|\eta\right|<2.5$~\footnote{Pseudorapidity is defined as $\eta=-\log{\tan{\theta \over 2}}$, where $\theta$ is the polar angle of the particle. }.  For the comparison with the Drell-Yan cross-sections, it is required that the invariant mass of the lepton pairs be in the range $71<m_{ll}<111\,\gev$. The cross-sections for the Drell-Yan process are calculated by requiring two leptons with $p_T>20\gev$ in the range $\left|\eta\right|<2.5$. It is required that the distance $\Delta R=\sqrt{(\Delta\phi)^2+(\Delta\eta)^2}$ between leptons be greater than $0.2$ and between the leptons and a parton be greater than $0.7$. In both calculations the impact of internal photon bremsstrahlung is not taken into account, as it is expected to be small in the inclusive quantities we study here.  

Table~\ref{tab:nominalzztoz_static} displays the cross-sections for the $ZZ$ and $Z^\ast$ at the LHC for different ranges of the invariant mass of the leptonic system, $m_{N\ell}$. The first column indicates the leptonic invariant mass bin for which the quantities in the subsequent columns are reported. The last column reports the ratio, ${ \sigma_{ZZ} \over \sigma_{Z^\ast}}$, which is defined as:
\begin{equation}
\label{eq:zztoz}
{ \sigma_{ZZ} \over \sigma_{Z^\ast}} = { \sigma_{q\overline{q}\rightarrow ZZ}^{NLO} + \sigma_{gg\rightarrow ZZ}^{LO} \over \sigma_{q\overline{q} \rightarrow Z^\ast}^{NLO} }
\end{equation}

The ratio defined above depends weakly on the invariant mass of the leptonic system. The increase of the ratio is partially induced by the enhancement of the relative contribution of the $gg\to ZZ$ process. In the first row the relative contribution of $gg\to ZZ$ to the total $ZZ$ cross-section is 13\,\% and it grows to 17\,\% in the last row.
%%% John 
This is linked to the choice of scale implemented here, that could be regarded as unnaturally small for large values of the leptonic invariant mass.
%%% John 
When the scales are chosen equal to the leptonic invariant mass the ratio ${ \sigma_{ZZ} \over \sigma_{Z^\ast}}$ is stable in the range $250<m_{N\ell}<1000\,\gev$ to better than 5\,\%.

%\begin{table}[ht]
%\caption{Cross-sections (in fb) for the $ZZ$ and $Z^*$ at the LHC for different ranges of the invariant mass of the leptonic system (in gev). The event selection specified in Section~\ref{sec:zz} is applied. \label{tab:nominalzztoz}}
%\begin{center}
%\begin{tabular}{|c|c|c|c|c|}
%\hline
%Mass Range &$\sigma_{q\overline{q}
%\rightarrow Z^*}^{NLO}$ & $\sigma_{q\overline{q}
%\rightarrow ZZ}^{NLO}$& $\sigma_{gg
%\rightarrow ZZ}^{LO}$&  ${ \sigma_{ZZ} \over \sigma_{Z^*}}\times 10^3$ \\
%\hline
%200 -  250  & 1800 &  8.14 & 0.631 &  4.87 \\ 
%250 -  300  &  758 &  3.70 & 0.274 &  5.24 \\ 
%300 -  350  &  376 &  1.84 & 0.124 &  5.22 \\ 
%350 -  400  &  205 &  1.01 & 0.062 &  5.23 \\ 
%400 -  450  &  120 &  0.60 & 0.037 &  5.34 \\ 
%450 -  500  &   75.1 &  0.38 & 0.023 &  5.39 \\ 
%500 -  750  &  132 &  0.67 & 0.041 &  5.35 \\ 
%750 - 1000  &   26.2 &  0.14 & 0.009 &  5.48 \\ 
%\hline
%\end{tabular}
%\end{center}
%\end{table}

\begin{table}[t]
\caption{Scale-driven uncertainties for the $ZZ$ and $Z^\ast$ production at the LHC for different ranges of the invariant mass of the leptonic system (in $\gev$). The first and second sub-columns give the cross-sections (in fb) and relative scale-driven uncertainties (in \%), respectively.  The first and second rows under each mass range correspond to the two extreme scale configurations (see text). The event selection specified in Section~\ref{sec:zz} is applied. Results are given for $\sqrt{s}=14\,\tev$. \label{tab:errorzztoz_static}}
\begin{center}
\begin{tabular}{|c||c|c||c|c||c|c||c|c|}
\hline
$m_{N\ell}$ range & \multicolumn{2}{c||}{$\sigma_{q\overline{q}
\rightarrow Z^\ast}^{NLO}$} & \multicolumn{2}{c||}{$\sigma_{q\overline{q}
\rightarrow ZZ}^{NLO}$} & \multicolumn{2}{c||}{$\sigma_{gg
\rightarrow ZZ}^{LO}$}
& \multicolumn{2}{c|}{${ \sigma_{ZZ} \over \sigma_{Z^\ast}}\cdot 10^3$} \\ \hline
200 -  250  &  929.4 &   4.8 &  4.17 &   4.3 &  0.96 &  62.0 &  5.52 &   6.6 \\ 
  &  793.4 & -10.5 &  3.57 & -10.6 &  0.38 & -36.4 &  4.98 &  -3.8 \\ 
250 -  300  &  396.0 &   5.2 &  1.93 &   5.9 &  0.42 &  57.3 &  6.06 &   9.3 \\ 
  &  341.9 &  -9.2 &  1.66 &  -9.0 &  0.18 & -33.9 &  5.32 &  -4.1 \\ 
300 -  350  &  195.2 &   4.9 &  0.98 &   5.5 &  0.20 &  60.0 &  6.15 &   8.7 \\ 
  &  170.4 &  -8.5 &  0.85 &  -8.5 &  0.08 & -34.5 &  5.45 &  -3.8 \\ 
350 -  400  &  108.5 &   5.6 &  0.55 &   3.4 &  0.11 &  64.1 &  6.08 &   4.3 \\ 
  &   97.7 &  -5.0 &  0.48 & -10.0 &  0.04 & -36.2 &  5.35 &  -8.3 \\ 
400 -  450  &   63.5 &   5.0 &  0.35 &  10.6 &  0.07 &  70.3 &  6.65 &  11.8 \\ 
  &   57.4 &  -5.1 &  0.30 &  -6.4 &  0.03 & -36.7 &  5.47 &  -8.0 \\ 
450 -  500  &   40.6 &   6.7 &  0.22 &  11.0 &  0.05 &  72.5 &  6.66 &  10.9 \\ 
  &   36.2 &  -4.8 &  0.19 &  -6.0 &  0.02 & -38.5 &  5.72 &  -4.7 \\ 
500 -  750  &   75.8 &   5.4 &  0.41 &  11.5 &  0.10 &  82.8 &  6.80 &  14.8 \\ 
  &   70.1 &  -2.6 &  0.34 &  -8.9 &  0.03 & -39.9 &  5.29 & -10.7 \\ 
750 - 1000  &   14.9 &   8.7 &  0.08 &   8.2 &  0.03 &  96.3 &  7.81 &  13.5 \\ 
  &   13.6 &  -0.5 &  0.07 &  -8.1 &  0.01 & -44.2 &  5.93 & -13.8 \\ 
\hline
\end{tabular}
\end{center}
\end{table}

Table~\ref{tab:errorzztoz_static} shows the scale-driven uncertainties for the quantities considered in Table~\ref{tab:nominalzztoz_static}. The first sub-columns report the maximum and minimum values of the cross-sections and the ratio ${ \sigma_{ZZ} \over \sigma_{Z^\ast}}$ for the scale choice stated above. The scale-driven error on ${ \sigma_{ZZ} \over \sigma_{Z^\ast}}$ is dominated by the scale-driven error of the $gg\to ZZ$ process. 

The higher order QCD corrections to the $gg\to ZZ$ process may be large and it is conceivable that they will not be known by the time of the analysis of the first data by the experiments at the LHC. A conservative estimate of the scale-driven error on ${ \sigma_{ZZ} \over \sigma_{Z^\ast}}$ could be obtained by multiplying $\sigma_{gg\rightarrow ZZ}^{LO}$ by a factor of two while keeping the same relative error as reported in Table~\ref{tab:errorzztoz_static}. This would yield a scale error of less than 15\,\% for ${ \sigma_{ZZ} \over \sigma_{Z^\ast}}$.

%%% John 
We have also performed a study of the errors on ${ \sigma_{q\overline{q}\rightarrow ZZ}^{NLO} \over \sigma_{q\overline{q}\rightarrow Z^{\ast}}^{NLO} } $ and ${ \sigma_{q\overline{q}\rightarrow ZZ}^{NLO} \over \sigma_{q\overline{q}\rightarrow Z}^{NLO} } $ due to the uncertainties in the parton densities, integrating over the mass of the leptonic system to find fractional deviations of 0.5\,\% and 1.4\,\% for $\sqrt{s}=14\,\tev$, respectively.
The $\sqrt{s}$ dependence of these errors is very weak. The fractional errors of the $gg\rightarrow ZZ$ cross-sections due to parton density uncertainties is of the order of $5\div 10~\%$ after integrating over the mass of the leptonic system, based on studies performed with NLO and LO parton density sets (see Ref.~\cite{Kauer:2004fg} for more details). A detailed study of the parton density error correlations of the ratios ${ \sigma_{ZZ} \over \sigma_Z}$  and ${ \sigma_{ZZ} \over \sigma_{Z^\ast}}$ is not performed here.
%%% John 

The experimental errors on ${ \sigma_{ZZ} \over \sigma_{Z^\ast}}$ will be dominated by the uncertainties on the lepton identification and isolation efficiencies. These effects cannot be estimated with our parton level study and we leave a systematic investigation of such effects using a parton shower to the experimental collaborations.
%This aspect of the prediction should be covered by the experiments. 

The QCD higher order corrections and the relative contribution from the $gg\to ZZ$ process to the total $ZZ$ cross-section depend little on the cuts on the lepton $p_T$. For instance, raising and lowering the lepton $p_T$ thresholds by $10\,\gev$ results in no significant change in the results. Therefore, the scale errors on the ratio ${ \sigma_{ZZ} \over \sigma_{Z^\ast}}$ reported in Table~\ref{tab:errorzztoz_static} should hold for similar configurations of leptonic cuts.
%%% John 
We have also investigated the stability of the ratio ${\sigma_{ZZ} \over \sigma_{Z^\ast}}$ with $\sqrt{s}$, in different bins of the mass of the leptonic system (Table~\ref{tab:ratiosdep}). The results are stable to better than 10\,\%. 
%%% John 
%This is an indication of the stablity of the ratio with respect to variations of the parton density in the proton.

%In principle, the QCD higher order corrections to the Drell-Yan process are known to NNLO~\cite{Hamberg:1990np}. The scale-driven uncertainties obtained with the NLO treatment are expected to be reduced when using the NNLO description~\cite{Anastasiou:2003ds,Melnikov:2006kv}. The scale-driven uncertainties on ${ \sigma_{ZZ} \over \sigma_{Z^\ast}}$ in Table~\ref{tab:errorzztoz_static} will be recomputed in a future update.

During the early stages of data taking, when the integrated luminosity will be  $O(100)\,$pb$^{-1}$, only a handful of events are expected to be reconstructed with four isolated leptons. For this scenario the accuracy reported in Table~\ref{tab:errorzztoz_static} is sufficient to establish evidence for a significant excess of four lepton events. However, as more data is accumulated a better accuracy will be required and in  Section~\ref{sec:ww} we propose the definition of another ratio that could help to further reduce the theoretical uncertainty.

\begin{table}[t]
\caption{Stability of the ratio ${\sigma_{ZZ} \over \sigma_{Z^\ast}}\cdot 10^3$ for different ranges of the invariant mass of the leptonic system (in $\gev$) as a function of the proton-proton collision center of mass energy (in \tev). }
\begin{center}
\begin{tabular}{|c||c|c|c|c|}
\hline 
%$\sqrt{s}$ & $200 - 250$ & $250 - 300$ &  $300 - 500$ & $ >500$ 
%\\ \hline
% 14  &  4.51 &  4.87 &  5.18 &  5.06
%\\ 
% 12  &  4.52 &  4.88 &  5.07 &  4.98
%\\ 
% 10  &  4.45 &  4.88 &  4.96 &  4.98
%\\ 
%  8  &  4.47 &  4.82 &  4.97 &  4.98
%\\ 
%  6  &  4.44 &  4.73 &  4.93 &  5.04
$\sqrt{s}$                                           & $200-250$ & $250-300$ &                                         $300-500$ & $
500-1000$ \\\hline
14  &  5.17 &  5.54 &  5.79 &  6.08
\\ \hline
10  &  4.98 &  5.33 &  5.48 &  5.61
\\ \hline
8  &  4.92 &  5.24 &  5.34 &  5.53
\\ \hline
\end{tabular}
\label{tab:ratiosdep}
\end{center}
\end{table}

So far we have considered the inclusive production of four leptons without further requirements on the final state. The production of $ZZ$ remains one of the leading backgrounds for searches of four leptons with moderate \met~\cite{Belyaev:20084l}. The production of missing transverse momentum in association with $ZZ$ is due to $\tau$ decays and instrumental $\met$. The latter is expected to be mainly due to the mismeasurement of hadronic activity recoiling against the $ZZ$ system and other factors.
%%% John 
The \met spectrum is driven to a large extent by the transverse momentum of the $ZZ$ system that,
in the same spirit as the inclusive cross sections, could be predicted by using the $p_T$ spectrum of the $Z^{\ast}$. A comparison of the two spectra has been made for $\sqrt{s}=14\,\tev$ with the help of the package RESBOS~\cite{RESBOS}, showing that they are very similar~\footnote{This package implements QCD NNLL/NLO order of calculation for $Z^{(\ast)}$~\cite{Ladinsky:1993zn,Balazs:1997xd,Landry:2002ix} and  QCD NLL/NLO order calculation for $ZZ$ production~\cite{Balazs:1998bm,Frederix:2008vb}}. This study suggests that the observed $\met$ spectrum in $Z^{\ast}$ events could indeed be used to predict the \met spectrum of $ZZ$ events. 
%%% John 

\subsection{{\boldmath $WW$} Production}
\label{sec:ww}

The hadronic production of $W$ pairs has been considered extensively in
the literature (for a review, see Ref.~\cite{Haywood:1999qg}). The NLO QCD 
corrections to $q\bar{q} \to WW$ have been presented in
Refs.~\cite{Ohnemus:1991kk,Frixione:1993yp}, while NLO calculations
for $q\bar{q} \to WW \to \ell\bar{\nu}\bar{\ell'}\nu'$ including spin
and decay angle correlations can be found in
Refs.~\cite{Ohnemus:1994ff,Dixon:1998py,Dixon:1999di,Campbell:1999ah,Grazzini:2005vw}.
%%% John 
Electro-weak corrections, which become
important at large $WW$ invariant masses, have been computed in Ref.~\cite{Accomando:2004de}.
The gluon-gluon induced contribution to on-shell $W$-pair production was first
calculated in Ref.~\cite{Glover:1988fe} for the case of massless
quarks circulating in the loop and then extended in
Ref.~\cite{Kao:1990tt} to include the top-bottom massive quark loop.
%%% John 
The first calculation of the gluon-induced
process $gg \to W^{\ast}W^{\ast} \to \ell\bar{\nu}\bar{\ell'}\nu'$,
including spin and decay angle correlations and allowing for arbitrary
invariant masses of the intermediate $W$ bosons in Ref.~\cite{Binoth:2006mf} is used here. 
%In a first step, and in order to quantify the general importance of the gluon-fusion process,
%we restrict our calculation to the leading contribution which arises
%from intermediate light quarks of the first two generations~\cite{Binoth:2006mf}.
We do not take into account gluon-gluon induced tree-level
processes of the type $gg \to WW q\bar q$, which have been found to be
strongly suppressed in hadronic $WZ$, $W\gamma$ and $Z\gamma$
production~\cite{Adamson:2002jb}.

\begin{table}[t]
\caption{Cross-sections for  $WW$ and $Z^{(\ast)}$ production and the ratios ${ \sigma_{WW} \over \sigma_Z}{\cdot 10^3}$ and ${ \sigma_{WW} \over \sigma_{Z^\ast}}$ (see text). The nominal cross-sections are given in the second row in fb except for $\sigma_{q\overline{q}\rightarrow Z}^{NLO}$, which is given in pb.  Cross-sections for  $Z^\ast$ are given for the range $M_{Z^\ast}>185\,\gev$. The event selection specified in Section~\ref{sec:ww} is applied. The fractional deviations (see text) in the second and third rows are given in $\%$. Results are given for $\sqrt{s}=14\,\tev$.\label{tab:nominalwwtoz_static}}
\begin{center}
\begin{tabular}{|c|c|c||c|c||c|c|}
\cline{2-7}
\multicolumn{1}{c|}{}&$\sigma_{q\overline{q}\rightarrow Z}^{NLO}$ &$\sigma_{q\overline{q}\rightarrow Z^\ast}^{NLO}$ & $\sigma_{q\overline{q}
\rightarrow WW}^{NLO}$& $\sigma_{gg
\rightarrow WW}^{LO}$& ${ \sigma_{WW} \over \sigma_Z}{\cdot 10^3}$ & ${ \sigma_{WW} \over \sigma_{Z^\ast}}$ \\
\hline
%Nom.& 1571 & 4513 & 1272 & 62.08 & 0.85  & 0.296\\ \hline
%Max. & -1.3 & +4.6 & +11.5 & +62.1  & +15.4  & +8.9   \\ \hline
%Min. & -15.7 &  -9.9 & -13.4 & -35.9 & +1.5 & -5.0 \\ \hline
%Nom. &  785.3 & 2256 &  636.0 &  31.04 &   0.85 &  0.296
%\\ \hline
%Max. &   6.2 &   4.6 &  11.5 &  62.1 &  16.1 &   9.4
%\\ \hline
%Min. & -15.7 &  -9.9 & -13.4 & -36.0 &  -8.6 &  -5.3
Nom. &  785.3 & 2256.4 &  636.0 &  31.04 &   0.85 &  0.296
\\ \hline
Max. &   6.2 &   4.6 &  11.5 &  62.1 &  16.1 &   9.4
\\ \hline
Min.  & -15.7 &  -9.9 & -13.4 & -36.0 &  -8.4 &  -5.3
\\ \hline
\end{tabular}
\end{center}
\end{table}

%%% John 
The package MCFM is used for the computation of the cross-sections due to $q\overline{q}\to WW \to \ell\bar{\nu}\bar{\ell'}\nu'$ in NLO QCD, $\sigma_{q\overline{q}\rightarrow WW}^{NLO}$~\cite{Campbell:1999ah} and we use the package {\tt gg2WW}, a parton-level integrator and event generator~\cite{Binoth:2006mf}, to compute the $gg \to W^{\ast}W^{\ast} \to \ell\bar{\nu}\bar{\ell'}\nu'$ cross-sections to leading order, $\sigma_{gg\rightarrow WW}^{LO}$.
As for $ZZ$ production, a complete NNLO QCD description of $WW$ production is not available, 
and we once again simply sum the cross-sections $\sigma_{q\overline{q}\rightarrow WW}^{NLO}$ and $\sigma_{gg\rightarrow WW}^{LO}$. 
%%% John 
%This will lead to conservative estimates of the theoretical errors.

We calculate the cross-sections requiring two opposite sign leptons with $p_T>20\,\gev$ in the range $\left|\eta\right|<2.5$ with the same $\Delta R$ requirements as in Section~\ref{sec:zz}. It is also required that the modulus of the vector sum of the transverse momenta of the neutrinos be greater than $20\,\gev$~\footnote{The sum of the transverse momentum of the neutrinos is not equivalent to the measured \met. These are a number of factors that, in addition to the detector smearing and other instrumental effects, distort the measurement: limited detector acceptance in $\eta$ and $p_T$, and neutrinos from fragmentation.}. The Drell-Yan cross-sections are obtained with the same cuts on the leptons as in Section~\ref{sec:zz}. The cross-section for the on-shell $Z$, $\sigma_{q\overline{q} \rightarrow Z}^{NLO}$, is obtained by requiring the invariant mass of the lepton pairs be in the range $71<m_{ll}<111\,\gev$ while the cross-section for an off-shell $Z$ corresponds to $m_{ll}>185\,\gev$.

Ratios analogous to Expression~\ref{eq:zztoz} are defined for $WW$ and $Z^{(*)}$ production, ${ \sigma_{WW} \over \sigma_Z}$  and ${ \sigma_{WW} \over \sigma_{Z^\ast}}$. Table~\ref{tab:nominalwwtoz_static} shows the cross-sections and scale-driven uncertainties for the $WW$ and $Z^{(\ast)}$ production together with the ratios ${ \sigma_{WW} \over \sigma_Z}$ and ${ \sigma_{WW} \over \sigma_{Z^\ast}}$. The maximum deviation for $\sigma_{q\overline{q} \rightarrow Z}^{NLO}$ is obtained when allowing both the renormalization and factorization scales to change at the same time, which is not the case for the rest of the processes considered here. The size of the error band for ${ \sigma_{WW} \over \sigma_Z}$  tends to be larger than that for ${ \sigma_{WW} \over \sigma_{Z^\ast}}$. 
%%% John 
Estimating higher order corrections by increasing the $gg\rightarrow WW$ cross-section by a factor of two enhances the scale-related fractional deviations of  ${ \sigma_{WW} \over \sigma_{Z^\ast}}$ and ${ \sigma_{WW} \over \sigma_Z}$, although they remain smaller than 15\,\% and 20\,\% respectively.
%%% John 

A detailed study of the errors due to the uncertainties in the parton densities is performed for ${ \sigma_{q\overline{q}\rightarrow WW}^{NLO} \over \sigma_{q\overline{q}\rightarrow Z}^{NLO} } $ and ${ \sigma_{q\overline{q}\rightarrow WW}^{NLO} \over \sigma_{q\overline{q}\rightarrow Z^{\ast}}^{NLO} } $ yielding fractional deviations of 1.3\,\% and 0.3\,\% for $\sqrt{s}=14\,\tev$, respectively. The $\sqrt{s}$ dependence of these errors is very weak. A detailed study of the parton densitiy error correlations of the ratios ${ \sigma_{WW} \over \sigma_Z}$  and ${ \sigma_{WW} \over \sigma_{Z^\ast}}$ is not performed here. Similarly as for the $gg\rightarrow ZZ$, the fractional errors of the $gg\rightarrow WW$ cross-sections due to parton density uncertainties is of the order of $5\div 10~\%$.

The cross-section ratios reported here correspond to an inclusive event selection, for which the relative contribution of the $gg\to WW$ diagrams to the total $WW$ cross-section is about 5\%. The relative contribution of the $gg\to WW$ process is significantly larger after the application of cuts for the Higgs boson search~[61, 62, 59].
%~\cite{Binoth:2005ua,Duhrssen:2005bz,Binoth:2006mf}. 
In addition to the expected large QCD higher order correction the scale error of the prediction of ${ \sigma_{WW} \over \sigma_Z}$  and ${ \sigma_{WW} \over \sigma_{Z^\ast}}$ will be significantly larger than reported in Table~\ref{tab:nominalwwtoz_static}. An alternative method for predicting the rate of the $WW$ background in the signal-like region for the Higgs boson search was suggested in Ref.~\cite{Aglietti:2006ne}.

%Although the kinematics of the decay products of the $gg\to WW$ and $q\overline{q}\to WW$ diagrams are different, it is very hard to disentangle the processes experimentally.  
The ratios ${ \sigma_{WW} \over \sigma_Z}$  and ${ \sigma_{WW} \over \sigma_{Z^\ast}}$ are sensitive to the contribution from the $gg\to WW$ diagrams. We strongly encourage the LHC experiments to measure these ratios in addition to the inclusive cross-section measurements. 

\begin{table}[t]
\caption{Cross-sections (in fb) and error due to scale variations for  $WW$ and $ZZ$ production and the ratio ${ \sigma_{WW} \over \sigma_{ZZ}}{\cdot 10^2}$  (see text). The event selections specified in Sections~\ref{sec:zz} and~\ref{sec:ww} are applied. The maximum and minimum fractional deviations (see tex) are given in $\%$. Results are given for $\sqrt{s}=14\,\tev$.\label{tab:ratiowwzz}}
%\caption{The dependence of the $WW$ and $ZZ$ cross-sections (in fb) and of the ratio ${\sigma(ZZ) \over \sigma(WW)}\cdot 10^{2}$ (see text) to on the center of mass energy of the proton-proton collision (in\,\tev).  \label{tab:ratiowwzz}}
\begin{center}
%\begin{tabular}{|c||c|c||c|c||c|c|}
\begin{tabular}{|c|c||c|c||c|c|}
\hline 
% $\sqrt{s}$
$\sigma_{WW}$ &                      $\delta\sigma_{WW}$ &                                            $\sigma_{ZZ}$ &                                                   $\delta\sigma_{ZZ}$ &                                            ${\sigma_{ZZ} \over \sigma_{WW}  } \cdot 10^2$ &                            $\delta{\sigma_{ZZ} \over \sigma_{WW}  }$ 
\\ \hline
  667.0 &  13.9 &  11.51 &  10.9 &  1.73 &   1.6
\\ 
  & -14.4 &  & -13.1 & &  -2.6
\\ \hline
%10 &  916.0 &   9.5 &  15.74 &   9.3 &  1.72 &   1.3
%\\ 
%  &  & -10.9 &  &  -9.7 & &  -0.2
%\\ \hline
%8 &  697.8 &  10.1 &  12.19 &   7.8 &  1.75 &   0.3
%\\ 
%  &  &  -7.8 &  &  -7.5 & &  -2.0
%\\ \hline
\end{tabular}
\end{center}
\end{table}

As anticipated in Section~\ref{sec:zz}, here we consider an additional ratio in order to further suppress the theoretical uncertainties of ${ \sigma_{ZZ} \over \sigma_{Z^\ast}}$ due to the error on  $\sigma_{gg\rightarrow ZZ}^{LO}$. We evaluate the scale related errors of:
\begin{equation}
{ \sigma_{ZZ} \over \sigma_{WW}} = { \sigma_{q\overline{q}\rightarrow ZZ}^{NLO} + \sigma_{gg\rightarrow ZZ}^{LO} \over \sigma_{q\overline{q}\rightarrow WW}^{NLO} + \sigma_{gg\rightarrow WW}^{LO} }
\end{equation}

Table~\ref{tab:ratiowwzz} displays the cross-sections $\sigma_{WW}=\sigma_{q\overline{q}\rightarrow WW}^{NLO} + \sigma_{gg\rightarrow WW}^{LO}$ and $\sigma_{ZZ}=\sigma_{q\overline{q}\rightarrow ZZ}^{NLO} + \sigma_{gg\rightarrow ZZ}^{LO}$ (in fb) together with the values of ${ \sigma_{ZZ} \over \sigma_{WW}}$ for $\sqrt{s}=14\,\tev$.  Results are computed after the application of the cuts specified in this Section and Section~\ref{sec:zz}.  The deviations of the cross sections and the ratio ${ \sigma_{ZZ} \over \sigma_{WW}}$ due to scale variations are given as percentages. The first and second numbers in the second, fourth and sixth columns correspond to the maximum and minimum deviations due to the scale variations, respectively.  The scale-driven uncertainties of ${ \sigma_{ZZ} \over \sigma_{WW}}$ are better than 5\,\%. It is relevant to note that the  central value of this ratio to NLO (ignoring the $gg\rightarrow VV$ processes) is equal to 1.58 and the maximum and minimum fractional deviations due to scale variations are 3.9\,\% and -6.7\,\%, respectively. The reduction of the scale-driven variations when adding the $gg\rightarrow VV$ process may be accidental. The results are also relatively stable with respect to potentially large QCD higher order corrections on the  $gg\rightarrow VV$ processes. For instance, an enhancement of the $gg\rightarrow VV$ cross-sections by a factor of two yields ${ \sigma_{ZZ} \over \sigma_{WW}}=1.86$, a 7.5\,\% deviation with respect to the nominal value.
%%% John 
We note also that the values of ${ \sigma_{ZZ} \over \sigma_{WW}}$ are stable with respect to $\sqrt{s}$ to better than 5\,\% and the error bands depend only weakly on $\sqrt{s}$.
%%% John 

The rate of di-leptons from $WW$ is expected to be much larger that that of four leptons from $ZZ$, indicating that the statistical error of a prediction made using this ratio will be negligible. However, the experimental errors on the rate of di-leptons with \met may be a dominating factor in the overall uncertainty of ${ \sigma_{ZZ} \over \sigma_{WW}}$. This aspect of the prediction needs to be evaluated by the experimentalists.

Here we have considered the production of opposite-sign leptons. The production of same-sign leptons is of great interest for searches of physics beyond the SM.
%%% John 
The process $qq\rightarrow W^\pm W^\pm qq$ (that we have not considered above) is one of the leading SM backgrounds, especially for large values of \met~\footnote{The production of same sign leptons form the $t\overline{t}$ decays dominates the production from SM processes of same-sign leptons at low \met.}. Nevertheless, a similar ratio ${ \sigma_{W^\pm W^\pm} \over \sigma_{Z^{(\ast)}}}$ could be defined, whose theoretical uncertainties would be dominated by the large error from unknown higher order corrections to the $qq\rightarrow W^\pm W^\pm qq$ cross-section~\cite{Kulesza:1999zh}.
%%% John 

\subsection{{\boldmath $ZW$} Production}
\label{sec:zw}

The hadronic production of $ZW$ is known to NLO~\cite{Ohnemus:1994ff,Dixon:1998py} including the leptonic decays. As opposed to the production of $ZZ$ and $WW$, $ZW$ production is not subject to gluon-gluon induced quark loop diagrams and, furthermore, the gluon induced $gg\rightarrow ZWq\overline{q}$ contributions are small~\cite{Adamson:2002jb}. 
%On the other hand, the production of $ZW$ bosons is not the result of $q\overline{q}$ annihilation.
% rendering the ratio of the inclusive $ZW$ cross-section to that of Drell-Yan more dependent on assumptions of parton densities in the proton.

The package MCFM is used here for the computation of the cross-sections due to $qq'\to Z^\ast(\gamma^\ast)W^\pm\to \ell\bar{\ell}\ell'\bar{\nu'}$ to QCD  NLO, $\sigma_{qq'\rightarrow ZW}^{NLO}$~\cite{Campbell:1999ah}. We compute the cross-sections for $ZW$ production requiring three leptons in the range $\left|\eta\right|<2.5$. For the leading lepton it is required that $p_T>20\,\gev$ and the sub-leading leptons have $p_T>10\,\gev$. The same $\Delta R$ cuts as those required in Section~\ref{sec:zz} are applied here. The invariant mass of the leptons from $Z^\ast(\gamma^\ast)$ is required to be larger than 20\,\gev. It is also required that the $p_T$ of the neutrino be larger than 20\,\gev. The cross-sections for $Z^\ast$ are obtained for $M_{Z^\ast}>195\,\gev$.

Table~\ref{tab:nominalzwtoz_static} displays the cross-sections for $ZW$ and $Z^{\ast}$ and the ratios ${ \sigma_{ZW} \over \sigma_Z}{\cdot 10^3}$ and ${ \sigma_{ZW} \over \sigma_{Z^\ast}}$. The cross-sections for $Z$ in Table~\ref{tab:nominalwwtoz_static} are used for the computation of the ratio ${ \sigma_{ZW} \over \sigma_Z}$. The maximum deviation of the ratio ${ \sigma_{ZW} \over \sigma_Z}$ occurs when scales are varied in the same direction. 
%As in the case of the ratio ${ \sigma_{WW} \over \sigma_Z}$ discussed in Section~\ref{sec:ww}, the maximum deviation of the $Z$ cross-section does not happen when the renormalization and factorization scales are changed in opposite directions. The fractional deviation for ${ \sigma_{ZW} \over \sigma_Z}$ is equal to -11.3\,\% if the renormalization and factorization scales are scaled up at the same time. This is due to the fact the $Z$ cross-section increases by 6.2\,\% while the cross-section for $ZW$ decreases by 5.8\,\%. 
The uncertainty due to the scale variations is at least twice as large for ${ \sigma_{ZW} \over \sigma_Z}$ than it is for ${ \sigma_{ZW} \over \sigma_{Z^\ast}}$ to NLO.

\begin{table}[t]
\caption{Cross-sections for  $ZW$ and $Z^{\ast}$ production and the ratios ${ \sigma_{ZW} \over \sigma_Z}{\cdot 10^3}$ and ${ \sigma_{ZW} \over \sigma_{Z^\ast}}$ (see text). The nominal cross-sections are given in the second row in fb.  Cross-sections for  $Z^\ast$ are given for the range $M_{Z^\ast}>195\,\gev$. The event selection specified in Section~\ref{sec:zw} is applied. The fractional deviations in the second and third rows are given in $\%$. Results are given for $\sqrt{s}=14\,\tev$. \label{tab:nominalzwtoz_static}}
\begin{center}
\begin{tabular}{|c|c|c||c|c|}
\cline{2-5}
\multicolumn{1}{c|}{}&$\sigma_{q\overline{q}\rightarrow Z^\ast}^{NLO}$ & $\sigma_{q\overline{q}
\rightarrow ZW}^{NLO}$& ${ \sigma_{ZW} \over \sigma_Z}{\cdot 10^3}$ & ${ \sigma_{ZW} \over \sigma_{Z^\ast}}$ \\
\hline
%Nominal & 3797 & 185.1 & 0.118  & 0.0487\\ \hline
%Maximum  & +4.6 & +12.9 & +14.4  & 7.9  \\ \hline
%Minimum  &  -9.2 & -12.0 & +4.5 & -3.0 \\ \hline
Nominal &  1898.4 &   92.5 &  0.118 & 0.0487
\\ \hline
Maximum &   4.6 &  12.9 &  16.0 &   7.9
\\ \hline
Minimum  &  -9.2 & -12.0 & -11.3 &  -6.5
\\ \hline
\end{tabular}
\end{center}
\end{table}

As in previous cases, the theoretical errors reported in Table~\ref{tab:nominalzwtoz_static} depend weakly on the event selection. The ratios presented in Table~\ref{tab:nominalzwtoz_static} have been evaluated for different values of the proton-proton center of mass energy in the range $6<\sqrt{s}<14\,\tev$. With respect to the values for $\sqrt{s}=14\,\tev$, the ratio ${ \sigma_{ZW} \over \sigma_Z}$ changes by about -8\,\% and -24\,\% for $\sqrt{s}=10$ and $6\,\tev$ respectively. The $\sqrt{s}$ dependence is milder for ${ \sigma_{ZW} \over \sigma_{Z^\ast}}$, with corresponding fractional deviations of $-4\,\%$ and $-10\,\%$. We have also studied the $\sqrt{s}$ dependence of these ratios in bins of the invariant mass of the $ZW$ system. For instance, in the $ZW$ mass bin $m_{ZW}>500\,\gev$ these fractional deviations for ${ \sigma_{ZW} \over \sigma_Z}$ are -15\,\% and -46\,\% at $\sqrt{s}=10$ and $6\,\tev$. The corresponding deviations for ${ \sigma_{ZW} \over \sigma_{Z^\ast}}$ (for which $m_{Z^\ast}>500\,\gev$ as well) are 3\,\% and 9\,\%, respectively. This is an indication that the ratio ${ \sigma_{ZW} \over \sigma_{Z^\ast}}$ is less prone to uncertainties related to parton densities in the proton. 	

A detailed study of the errors due to the uncertainties of the parton densities is performed for ${ \sigma_{ZW} \over \sigma_Z}$ and ${ \sigma_{ZW} \over \sigma_Z^{\ast}}$ yielding fractional deviations of 1.4\,\% and 0.4\,\% for $\sqrt{s}=14\,\tev$, respectively. The $\sqrt{s}$ dependence of these errors is very weak. 

\section{Jet Veto Survival Probability}
\label{sec:jvsp}

As pointed out in Section~\ref{sec:intro}, the requirement of a jet veto is instrumental in separating the multi-lepton production coming from the decays of $VV$ from that of $t\overline{t}$ production. The latter is associated with large hadronic activity and it is strongly suppressed by the application of a veto on high $p_T$ hadronic jets~\cite{CMSPTDR,LHCC99-14,Aad:2009wy}. 

The jet veto survival probability (JVSP, or $\epsilon_{jv}$) is defined as the fraction of the events with leptons passing the analysis requirements that do not display a parton, a quark or a gluon,  with a $p_T$ above a certain threshold in the range $\left|\eta\right|<5$~\footnote{The application of a jet veto requirement is usually referred to as the full jet veto. This is done to distinguish it from the application of a veto on jets in addition to two well separated jets, usually performed to isolate the Higgs boson produced via Vector Boson Fusion. The latter is not considered in this paper.}.
%%% John 
Here we attempt to address the possibility of predicting the JVSP for vector boson pairs by using the production of di-leptons from the Drell-Yan mechanism in a similar invariant mass range.
%%% John 

%%% John 
The nominal results for the JVSP are obtained with the QCD NLO ME used in previous sections. No corrections due to detector and hadronization effects are taken into account in the nominal results reported here, although one might expect them to mostly cancel out in the ratio defined here.
%%% John 
The impact of multiple gluon radiation will be discussed below. The results in this Section are obtained with the same settings and event selections as those used in Section~\ref{sec:inclusive}.

Tables~\ref{tab:cjv14} and~\ref{tab:cjvzw14} report results for the JVSP for the $WW$ and $ZW$ production. The central values, $\epsilon^{Z^{\ast}}_{jv}$, $\epsilon^{WW}_{jv}$ and $\epsilon^{ZW}_{jv}$ are reported, together with the maximum and minimum fractional deviations due to the scale variations, expressed as percentages. Results are reported for different values of the parton $p_T$ threshold (in $\gev$). Results are also shown for the ratios ${\epsilon^{WW}_{jv} \over \epsilon^{Z^{\ast}}_{jv}  }$ and ${\epsilon^{ZW}_{jv} \over \epsilon^{Z^{\ast}}_{jv}  }$, quoted for $M_{Z^\ast}>185\,\gev$ and $M_{Z^\ast}>195\,\gev$, respectively. The maximum deviations reported in Tables~\ref{tab:cjv14} and~\ref{tab:cjvzw14} for the ratios ${\epsilon^{WW}_{jv} \over \epsilon^{Z^{\ast}}_{jv}  }$ and ${\epsilon^{ZW}_{jv} \over \epsilon^{Z^{\ast}}_{jv}  }$ are less than 10\,\% and have a tendency to dececrease with increasing $p_T$ threshold. 
A more precise prediction of this quantity would require the use of the calculation of Drell-Yan and $VV$ production at NNLO. Only the former is available~\cite{Hamberg:1990np,Anastasiou:2003ds,Melnikov:2006kv,Catani:2009sm}. We could note that some steps toward this have already been taken, such as NLO calculations of WW+jet production~\cite{Dittmaier:2007th,Campbell:2007ev,Sanguinetti:2008xt}.

\begin{table}[t]
\caption{Central values and scale-driven uncertainties of the jet veto survival probability for the $Z^{\ast}$ and $WW$ production for different values of the parton $p_T$ threshold (in $\gev$). For a veto the parton is required to be the range $\left|\eta\right|<5$. Results for  $Z^\ast$ are given for the range $M_{Z^\ast}>185\,\gev$. The scale related uncertainty is expressed in $\%$. Results are obtained for $\sqrt{s}=14\,\tev$. \label{tab:cjv14}}
\begin{center}
\begin{tabular}{|c||c|c||c|c||c|c|}
\hline 
 $p_{T}$ & $\epsilon^{Z^{\ast}}_{jv}$ &            $\delta\epsilon^{Z^{\ast}}_{jv}$ &                                  $\epsilon^{WW}_{jv}$ &                                                 $\delta\epsilon^{WW}_{jv}$ &                                          ${\epsilon^{WW}_{jv} \over \epsilon^{Z^{\ast}}_{jv}  }$ &                $\delta{\epsilon^{WW}_{jv} \over \epsilon^{Z^{\ast}}_{jv}  }$ 
\\ \hline
 20  &  0.67 &   8.5 &  0.52 &  11.9 &  0.78 &   5.1
\\ 
  &  & -13.2 &  & -15.2 & &  -3.2
\\ \hline
% 25  &  0.72 &   6.4 &  0.58 &   9.6 &  0.81 &   4.0
%\\ 
%  &  &  -9.9 &  & -11.8 & &  -2.9
%\\ \hline
 30  &  0.76 &   5.1 &  0.63 &   8.3 &  0.82 &   3.6
\\ 
  &  &  -7.8 &  &  -9.1 & &  -2.1
\\ \hline
% 35  &  0.79 &   4.1 &  0.67 &   7.4 &  0.84 &   3.3
%\\ 
%  &  &  -6.3 &  &  -7.3 & &  -2.1
%\\ \hline
 40  &  0.82 &   3.5 &  0.70 &   6.6 &  0.85 &   3.0
\\ 
  &  &  -5.3 &  &  -5.9 & &  -1.9
\\ \hline
% 45  &  0.84 &   3.2 &  0.72 &   6.0 &  0.86 &   2.8
%\\ 
%  &  &  -4.4 &  &  -5.4 & &  -1.8
%\\ \hline
 50  &  0.86 &   2.9 &  0.75 &   5.5 &  0.87 &   2.6
\\ 
  &  &  -3.8 &  &  -5.0 & &  -1.8
\\ \hline
% 55  &  0.87 &   2.6 &  0.77 &   5.1 &  0.88 &   2.5
%\\ 
%  &  &  -3.3 &  &  -4.7 & &  -1.7
%\\ \hline
% 60  &  0.88 &   2.4 &  0.79 &   4.7 &  0.89 &   2.3
%\\ 
%  &  &  -2.9 &  &  -4.3 & &  -1.6
%\\ \hline
% 65  &  0.89 &   2.1 &  0.80 &   4.3 &  0.90 &   2.1
%\\ 
%  &  &  -2.6 &  &  -4.0 & &  -1.4
%\\ \hline
% 70  &  0.90 &   1.9 &  0.81 &   4.1 &  0.90 &   2.1
%\\ 
%  &  &  -2.4 &  &  -3.7 & &  -1.3
%\\ \hline
% 75  &  0.91 &   1.8 &  0.83 &   3.7 &  0.91 &   1.9
%\\ 
%  &  &  -2.2 &  &  -3.3 & &  -1.1
%\\ \hline
% 80  &  0.92 &   1.6 &  0.84 &   3.4 &  0.91 &   1.7
%\\ 
%  &  &  -2.0 &  &  -3.2 & &  -1.1
%\\ \hline
% 85  &  0.93 &   1.5 &  0.85 &   3.1 &  0.92 &   1.6
%\\ 
%  &  &  -1.9 &  &  -3.1 & &  -1.2
%\\ \hline
% 90  &  0.93 &   1.4 &  0.86 &   3.0 &  0.92 &   1.5
%\\ 
%  &  &  -1.8 &  &  -2.9 & &  -1.1
%\\ \hline
% 95  &  0.94 &   1.3 &  0.87 &   2.7 &  0.93 &   1.4
%\\ 
%  &  &  -1.7 &  &  -2.7 & &  -1.1
%\\ \hline
100  &  0.94 &   1.2 &  0.88 &   2.6 &  0.93 &   1.3
\\ 
  &  &  -1.6 &  &  -2.6 & &  -1.1
\\ \hline
\end{tabular}
\end{center}
\end{table}

The impact on $\epsilon^{Z^{\ast}}_{jv}$ of final state radiation, hadronization and multiple gluon radiation (by means of the parton shower approximation) are studied with Pythia~\cite{Sjostrand:2000wi,Sjostrand:2001yu}. The impact of these effects on $\epsilon^{WW}_{jv}$ are studied with the MC@NLO~\cite{Frixione:2002bd} and ALPGEN~\cite{Mangano:2002ea} packages. It is observed that the effect of multiple gluon radiation on the JVSP for parton $p_T$ threshold values in the range $20<p_T<30\,\gev$ is significant. After taking account of the hadron to parton corrections it is observed that the JVSP decreases by about 10\,\% for parton $p_T$ thresholds around $30\,\gev$. This effect diminishes as the $p_T$ threshold increases. The impact of this effect on the ratio ${\epsilon^{WW}_{jv} \over \epsilon^{Z^{\ast}}_{jv}  }$ is less than 5\,\% for the same parton $p_T$ threshold, although more detailed studies are required in order to determine the theoretical errors on this correction. 

The JVSP is also evaluated for $Z$ events, $\epsilon^{Z}_{jv}$. The values of $\epsilon^{Z}_{jv}$ are 13\,\% and 12\,\% greater than $\epsilon^{Z^{\ast}}_{jv}$ for the values of the parton $p_T$ theresholds of 20 and 30\,\gev, respecively. The use of off-shell $Z$ events is preferred to predict the JVSP of $VV$. The use of on-shell $Z$ events for these studies is not precluded, although it would lead to enhanced theoretical errors on the ratios discussed here.

%%% John 
The values of $\epsilon^{WW}_{jv}$ reported here do not include the contribution from the gluon-gluon initiated processes discussed above since they do not include any radiation upon which to veto.
A calculation of the rate of $gg\rightarrow WWj$ would therefore greatly improve the estimates of the quantity $\epsilon^{WW}_{jv}$ given here
%%% John 
Assuming that the relative rate $gg\rightarrow WWj$ with respect to the total $WW$ rate is double that reported in Table~\ref{tab:nominalwwtoz_static}, the maximum and minimum possible fractional deviations of $\epsilon^{WW}_{jv}$ would be about 5\,\% and -9\,\% for the parton $p_T$ thereshold of 30\,\gev, respectively. These correspond to unphysical extreme cases when the JVSP for the $gg\rightarrow WW$ process is assumed to be 100 and 0\,\%, respectively. It is important to note that these statements are applicable only to event selections similar to the ones chosen in this paper.

It is important to note that selecting off-shell $Z$ bosons in association with at least one high $p_T$ jet requires to subtract $t\overline{t}$ backgrounds. The latter can be suppressed by the application of an \met cut not pointing in the direction of the lead jet in the event. Further studies are required to address the contamination of $t\overline{t}$ backgrounds. In order to circumvent this issue the ratio of the expected rate for Drell-Yan events to that of $VV$ after a jet veto can be defined. The errors on these ratios are similar to those of the ratios reported in this Section.

\begin{table}[t]
\caption{Central values and scale-driven uncertainty of the jet veto survival probability for the $Z^{\ast}$ and $ZW$ production for different values of the parton $p_T$ threshold (in $\gev$). For a veto the parton is required to be the range $\left|\eta\right|<5$. Results for  $Z^\ast$ are given for the range $M_{Z^\ast}>195\,\gev$. The scale-driven uncertainty is expressed in $\%$. Results are obtained for $\sqrt{s}=14\,\tev$. \label{tab:cjvzw14}}
\begin{center}
\begin{tabular}{|c||c|c||c|c||c|c|}
\hline 
 $p_{T}$ & $\epsilon^{Z^{\ast}}_{jv}$ &            $\delta\epsilon^{Z^{\ast}}_{jv}$ &                                  $\epsilon^{ZW}_{jv}$ &                                                 $\delta\epsilon^{ZW}_{jv}$ &                                          ${\epsilon^{ZW}_{jv} \over \epsilon^{Z^{\ast}}_{jv}  }$ &                $\delta{\epsilon^{ZW}_{jv} \over \epsilon^{Z^{\ast}}_{jv}  }$ 
\\ \hline
 20  &  0.67 &   8.5 &  0.48 &  13.2 &  0.71 &   6.3
\\ 
  &  & -13.2 &  & -15.3 & &  -7.3
\\ \hline
% 25  &  0.72 &   6.4 &  0.53 &  11.1 &  0.73 &   5.5
%\\ 
%  &  &  -9.9 &  & -12.2 & &  -6.4
%\\ \hline
 30  &  0.76 &   5.1 &  0.57 &   9.7 &  0.75 &   5.0
\\ 
  &  &  -7.8 &  & -10.8 & &  -5.9
\\ \hline
% 35  &  0.79 &   4.1 &  0.61 &   8.7 &  0.76 &   4.6
%\\ 
%  &  &  -6.3 &  &  -9.9 & &  -5.5
%\\ \hline
 40  &  0.82 &   3.5 &  0.64 &   7.7 &  0.78 &   4.0
\\ 
  &  &  -5.3 &  &  -9.2 & &  -5.3
\\ \hline
% 45  &  0.84 &   3.2 &  0.66 &   7.1 &  0.79 &   3.8
%\\ 
%  &  &  -4.4 &  &  -8.5 & &  -5.0
%\\ \hline
 50  &  0.86 &   2.9 &  0.68 &   6.5 &  0.80 &   3.5
\\ 
  &  &  -3.8 &  &  -7.8 & &  -4.7
\\ \hline
% 55  &  0.87 &   2.6 &  0.70 &   6.1 &  0.81 &   3.4
%\\ 
%  &  &  -3.3 &  &  -7.5 & &  -4.6
%\\ \hline
% 60  &  0.88 &   2.4 &  0.72 &   5.6 &  0.82 &   3.2
%\\ 
%  &  &  -2.9 &  &  -7.0 & &  -4.4
%\\ \hline
% 65  &  0.89 &   2.1 &  0.74 &   5.4 &  0.83 &   3.2
%\\ 
%  &  &  -2.6 &  &  -6.6 & &  -4.1
%\\ \hline
% 70  &  0.90 &   1.9 &  0.75 &   5.0 &  0.83 &   3.0
%\\ 
%  &  &  -2.4 &  &  -6.3 & &  -4.0
%\\ \hline
% 75  &  0.91 &   1.8 &  0.77 &   4.7 &  0.84 &   2.9
%\\ 
%  &  &  -2.2 &  &  -6.1 & &  -4.0
%\\ \hline
% 80  &  0.92 &   1.6 &  0.78 &   4.4 &  0.85 &   2.7
%\\ 
%  &  &  -2.0 &  &  -5.8 & &  -3.8
%\\ \hline
% 85  &  0.93 &   1.5 &  0.79 &   4.2 &  0.86 &   2.6
%\\ 
%  &  &  -1.9 &  &  -5.6 & &  -3.7
%\\ \hline
% 90  &  0.93 &   1.4 &  0.80 &   4.0 &  0.86 &   2.5
%\\ 
%  &  &  -1.8 &  &  -5.2 & &  -3.5
%\\ \hline
% 95  &  0.94 &   1.3 &  0.81 &   3.8 &  0.87 &   2.4
%\\ 
%  &  &  -1.7 &  &  -4.9 & &  -3.2
%\\ \hline
100  &  0.94 &   1.2 &  0.82 &   3.6 &  0.87 &   2.3
\\ 
  &  &  -1.6 &  &  -4.6 & &  -3.0
\\ \hline
\end{tabular}
\label{}
\end{center}
\end{table}

The ability to control the JVSP for $VV$ production gives us a powerful handle to understand better the interplay between this and related $t\overline{t}$ backgrounds. The residual contribution of $t\overline{t}$ events after the application of a jet veto can be evaluated by extrapolation after subtracting for the contribution of $VV$ production. This technique and the study of the corresponding theoretical errors will be developed further in the future. 

A change in the $\left|\eta\right|$ bound from 5 to 3 does not change the JVSP of $VV$ considerably, as most of the partons lie in the range $\left|\eta\right|<3$. The reconstruction of hadronic jets in the range $3<\left|\eta\right|<5$ is more challenging than in the central regions of the detector and we therefore consider the possibility of relaxing the requirement $\left|\eta\right|<5$ used by the experiments. This would also be appropriate for the early stages of data taking.
%%% John 
The LO matrix elements for $t\overline{t}$ ($0j$) and $t\overline{t}j$  available in MCFM are used for the evaluation of the JVSP for the $t\overline{t}$  processes, using the event selection described in Section~\ref{sec:ww}.
%%% John 
The JVSP for the $t\overline{t}$ ($0j$) and $t\overline{t}j$ production increases by 26\,\% and 56\,\% for a parton $p_T$ threshold of $30\,\gev$~\footnote{It is important to note that the JVSP for the $t\overline{t}j$ production is less than half of that of $t\overline{t}$ ($0j$) production, indicating that the $t\overline{t}j$ production will play a minor role when considering event selections with a tight full jet veto. This is the case of the Higgs boson search with the $H\rightarrow WW$ decay. }. These results represent a mild increase of the overall background contributions in analyses with a tight jet veto.

The $\sqrt{s}$ dependence of the main hadronic variables and the JVSP are studied for $WW$ and $Z^{\ast}$ production.  Table~\ref{tab:sdepjet} displays the  $\sqrt{s}$ evolution of the average $p_T$ (in $\gev$) and $\left|\eta\right|$ of the parton. As expected from the perturbative analysis, the $p_T$ of the parton decreases with decreasing $\sqrt{s}$. As $\sqrt{s}$ decreases the longitudinal boost of particles produced in the hard scattering of the proton-proton collision decreases, which is reflected by the decrease of the average parton $\left|\eta\right|$. The increase of the average parton $p_T$ and decrease of the parton $\left|\eta\right|$ as $\sqrt{s}$ decreases are two competing effects as far as the JVSP is concerned. Table~\ref{tab:sdepjet} shows $\epsilon_{jv}^{WW}$, $\epsilon_{jv}^{Z^\ast}$ and ${\epsilon_{jv}^{WW} \over \epsilon_{jv}^{Z^{\ast}}}$ (calculated for a $p_T$ threshold of 30\,\gev) as a function of $\sqrt{s}$. Overall, the JVSP increases mildly as $\sqrt{s}$ decreases. The ratio ${\epsilon_{jv}^{WW} \over \epsilon_{jv}^{Z^{\ast}}}$ is even more stable, varying by less than 10\,\% in the rate $6<\sqrt{s}<14\,\tev$. The same discussion applies to the JVSP for the $ZW$ production. The errors reported in Tables~\ref{tab:cjv14} and~\ref{tab:cjvzw14} have also been studied as a function of $\sqrt{s}$. The errors due to scale variations in ${\epsilon^{WW}_{jv} \over \epsilon^{Z^{\ast}}_{jv}  }$ and ${\epsilon^{ZW}_{jv} \over \epsilon^{Z^{\ast}}_{jv}  }$ have a tendency to increase, while remaning smaller than 10\,\%. The JVSP increases by 8\,\% and 4\,\% for $WW$ and $Z^{\ast}$, respectively, when going from $\sqrt{s}=14\,\tev$ to $\sqrt{s}=10\,\tev$. The ratio ${\epsilon_{jv}^{WW} \over \epsilon_{jv}^{Z^{\ast}}}$ is even more stable with $\sqrt{s}$.

As shown in Table~\ref{tab:sdepjet}, the fraction of partons in the range $\left|\eta\right|<3$ increases as $\sqrt{s}$ decreases.  This effect is stronger when considering the $\eta$ distributions of the decay products of the $t\overline{t}$ ($0j$) and $t\overline{t}j$ production. For instance, the JVSP increases by 18\,\% and 8\,\% for the $t\overline{t}$ ($0j$) production when relaxing the jet veto $\eta$ range (for a parton $p_T$ threshold of $30\,\gev$) for $\sqrt{s}=10$ and  $6\,\tev$, respectively. This should be compared with 26\,\% for $\sqrt{s}=14\,\tev$. This further  motivates relaxing the $\left|\eta\right|$ range of the jet veto, especially during the early stages of data taking. 

\begin{table}[t]
\caption{The dependence of various jet related variables of $WW$ and $Z^{\ast}$ production on the center of mass energy of proton-proton collision (in \tev). Results for the central jet veto survival probability are given for a $p_T$ threshold of 30\,\gev. \label{tab:sdepjet}}
\begin{center}
\begin{tabular}{|c||c|c|c||c|c|c||c|}
\cline{2-7}
\multicolumn{1}{c}{}&  \multicolumn{3}{|c||}{$WW$} & \multicolumn{3}{c|}{$Z^\ast$} & \multicolumn{1}{c}{} \\ \hline
$\sqrt{s}$ &                                                 $<p_{T}>$ & $<\left|\eta\right|>$ &                           $\epsilon_{jv}$ &                                                $<p_{T}>$ & $<\left|\eta\right|>$ &                           $\epsilon_{jv}$ &                                                ${\epsilon_{jv}^{WW} \over \epsilon_{jv}^{Z^{\ast}}}$             
\\ \hline
% 14  &  38.6 &  0.76 &  0.64 &  22.3 &  0.58 &  0.77 &  0.83
%\\ \hline
% 12  &  34.8 &  0.68 &  0.67 &  20.7 &  0.54 &  0.78 &  0.86
%\\ \hline
% 10  &  32.1 &  0.66 &  0.69 &  18.9 &  0.50 &  0.80 &  0.86
%\\ \hline
%  8  &  27.7 &  0.59 &  0.72 &  17.2 &  0.47 &  0.81 &  0.89
%\\ \hline
%  6  &  22.7 &  0.51 &  0.76 &  14.3 &  0.40 &  0.84 &  0.90
 14  &  42.0 &  0.78 &  0.63 &  23.9 &  0.58 &  0.76 &  0.82
\\ \hline
 10  &  34.6 &  0.68 &  0.68 &  21.0 &  0.53 &  0.78 &  0.86
\\ \hline
 8  &  30.1 &  0.62 &  0.71 &  18.1 &  0.47 &  0.81 &  0.87
\\ \hline
\end{tabular}

\end{center}
\end{table}

We have also studied the JVSP for $ZZ$ production. The rate of production of four leptons inclusively from SM processes is affected little by the decays of $t\overline{t}$. However, $t\overline{t}$, $t\overline{t}t\overline{t}$ and $t\overline{t}Z^\ast(\gamma^\ast)$ production contribute considerably in final states with large \met~\cite{Belyaev:20084l}. The application of a jet veto could be a viable option to suppress these backgrounds, for which it is necessary to understand the JVSP for $ZZ$ production. The JVSP for $ZZ$ production is considerably closer to $\epsilon^{Z^{\ast}}_{jv}$ than that of $WW$ production. The ratio ${\epsilon^{ZZ}_{jv} \over \epsilon^{Z^{\ast}}_{jv}}$ is equal to 0.92, 0.93 and 0.95 for $\sqrt{s}=14, 10$ and  $8\,\tev$, respectively. The errors due to the scale variations are very similar to those reported in Table~\ref{tab:cjv14}.

\section{Conclusions}
The search for new physics in final states with multiple charged leptons and \met carried by neutrinos or other particles escaping detection is arguably one of most the interesting among the feasible signatures at the LHC.
The production of two weak bosons  will be one of the most important sources of SM backgrounds for these final states. In this paper we consider several quantities that can help normalize rates for the production of weak boson pairs. Ratios involving the production of two weak bosons and Drell-Yan are investigated and the corresponding theoretical errors are evaluated. We consider ratios of inclusive cross-sections of $VV$ to that of the Drell-Yan process. We include the production of $WW$ and $ZZ$ through gluon-gluon fusion 
at order ${\cal O}(\alpha_s^2)$. We have considered the use of both on-shell and off-shell $Z$ production. We find that the use of off-shell $Z$ production tends to result in smaller errors due to scale variations and parton density uncertainties, depending on the ratio. We also consider the ratio of the cross-section for $ZZ$  to that for $WW$ production as an additional handle to reduce the theoretical errors on the prediction of the $ZZ$ cross-section. We strongly encourage the CMS and ATLAS experiments to  measure the ratio of the $WW$ cross-section to that of the $Z^{(\ast)}$ production. %This ratio would be sensitive to an anomalously large contribution from the $gg\rightarrow WW$ process.

The possibility of predicting the jet veto survival probability of $VV$ production with Drell-Yan is also considered. The use of off-shell $Z$ events is preferred for the prediction of the JVSP of $VV$ production. Our studies indicate that reducing the $\eta$ range of the jet veto used to suppress $t\overline{t}$ backgrounds is well motivated.  

Overall, the theoretical errors on the quantities presented here are less than $5\div 20~\%$.
%%% John 
Moreover, their dependence on the center of mass energy of the proton-proton collision is weak, so that early measurements at lower energies may help guide later more detailed studies.
%%% John 

\begin{acknowledgments}
We would like to thank S.~Dawson, G.~Dissertori, M.~Dittmar, P.~Fileviez-Perez, T.~Han, J.~Huston, B.~Mukhopadhyaya, F.~Petriello, J.~Qian, W.~Quayle, D.~Rebuzzi and D.~Zeppenfeld for most useful comments and discussions. N.~K. thanks the Higher Education Funding Council for England and the Science and Technology Facilities Council for financial support under the SEPnet Initiative. This work was supported in part by the DOE Grant No. DE-FG0295-ER40896 and under Task TeV of contract DE-FGO3-96-ER40956. The work of B.~M. is also supported by the Wisconsin Alumni Research Foundation.
\end{acknowledgments}

\bibliography{vbf,mycites}
\end{document}